\documentclass{elsart}
 
\usepackage{graphicx}
   
\usepackage{amssymb}
    
\begin{document}
\begin{frontmatter}
 
  
   
\title{Study of the influence of  surface anisotropy and
        lattice structure on the behaviour of a small magnetic cluster}
\author{Laura Hern\'{a}ndez \and Claire Pinettes
}                     
\address{Laboratoire de
Physique Th\'{e}orique et Mod\'elisation / CNRS-UMR 8089,\\
Universit\'{e} de Cergy-Pontoise, 
5 mail Gay Lussac, Neuville-sur-Oise, 
95031 Cergy-Pontoise, Cedex, France. }
%
%
\begin{abstract}
We have studied  by Monte Carlo simulations   the thermal behaviour of a small ($N=13$ particles)
cluster described by a Heisenberg model, including nearest-neighbour ferromagnetic interactions
and
radial surface anisotropy, in an applied magnetic field. We have studied three different lattice
structures: hexagonal close packed, face centered cubic and icosahedral.
We show that the  zero-field thermal behaviour depends not only on the value of the 
anisotropy constant
 but also on the  lattice structure.
The behaviour in an applied field, additionally depends,
on the different orientations of the field with respect
to the crystal axes. 
According to these relative  orientations,
hysteresis cycles show different  step-like characteristics. 
\end{abstract}
\end{frontmatter}  

%
\section{Introduction}
\label{intro}

The magnetic properties of clusters have been studied intensely for about two decades.
Surface and finite-size effects appear to play a key r\^ole for clusters containing
up to several hundred atoms.
An enhanced magnetic anisotropy at the surface has been observed for fine metallic 
particles~\cite{luis} and predicted by theoretical calculations
~\cite{hendriksen,pastor}.
One source for this anisotropy is the difference between the  coordination number at the surface  
and  in the bulk, inducing a change in the crystal 
fields~\cite{hendriksen,kodama,aharony,garanin}.
It has been experimentally found that this enhancement increases as the size of the particle 
decreases~\cite{luis}. 

Magnetic measurements are usually carried out by passing size selected clusters 
through a non uniform magnetic field as in the Stern-Gerlach experiment.
In these experiments the cluster size is known but the crystal structure 
as well as the initial orientation of the crystal axes of the cluster with respect 
to the field gradient direction are not.

A great deal of effort has been devoted to the theoretical and experimental 
study of transition metal (TM) clusters~\cite{bucher1,billas,meriko,bucher2} which are known to have a relatively low anisotropy energy  in the bulk
 compared to their exchange energy~\cite{hendriksen}.
For TM clusters, it has been shown that it is the localised character of the 3d 
electrons at the surface which enhances the surface anisotropy with respect to 
the volume anisotropy~\cite{pastor}. 
On the contrary very little is known for
 rare earth (RE) clusters, where the anisotropy energy is expected to be higher. Experimental 
 results on RE clusters show an anomalous behaviour (the broadening of the deflection profile 
 in a Stern-Gerlach  experiment at a finite temperature) which is absent in the case of TM 
 clusters~\cite{re1,gerion}.

Several studies which consider anisotropy terms have been performed,
 mostly concerning large clusters ($N \geq 1000$ atoms)
cut to regular shapes (spherical or ellipsoidal)
out of simple-cubic or spinel lattices. 
For these systems the competition between the surface anisotropy and 
the exchange term has been analysed as a function of temperature
~\cite{kach1,kach2,labaye}. 
The influence of surface effects on the hysteresis cycle at zero temperature
has also been shown~\cite{kach4}. 
Dimitrov and Wysin~\cite{dim1} have studied the zero-temperature behaviour 
of large particles with ferromagnetically coupled Heisenberg classical 
spins with either uniaxial random anisotropy or radial surface anisotropy.
They have studied the two-(three-)dimensional simple-cubic lattice  
with the cluster having either circular or rectangular (spherical or cubic) geometries. 
They find a step-like hysteresis cycle which is identified as a surface effect.  
This is also observed in their zero temperature study of spherical particles 
cut out of an fcc lattice structure~\cite{dim2}. 
To  have a  complete picture of these systems, it is of interest to investigate 
the thermal magnetic behaviour  
of small clusters, including an  anisotropy energy term.

In this work we  study a  simple classical Heisenberg model 
for an $N=13$ cluster with radial surface anisotropy, at finite temperature 
and in an applied external field, in  order to understand how the thermal magnetic
behaviour of the cluster is affected
by the existence of a  surface anisotropy term.
We will show how the interplay between the surface  anisotropy and the crystal 
structure, as well as the relative orientation of the applied
field with respect to the crystal axes, influence the magnetic behaviour of the cluster.

This article is organized as follows: in Section~\ref{sec:1}  we describe the studied
systems; in  Section~\ref{sec:2} we give the technical details of our calculation,
and in  Section~\ref{sec:3} we present the  results. Finally, in  Section~\ref{sec:4}, 
we summarise and discuss our results.

\section{Description of the systems studied}
\label{sec:1}

Among the small amount of information concerning RE clusters, reference~\cite{re4}
shows, by ab-initio 
electronic  structure calculations that  $Gd_{13}$ clusters adopt  a hexagonal closed
packed (hcp) crystal structure.
Numerical simulation results of cluster agregation starting from $N=13$ 
separated atoms placed in either  Lennard-Jones or  Gupta potentials
yield very similar cohesion energies for  hcp, fcc  and 
icosahedral lattices~\cite{sugano}.  
These leads us to  base our  study on   $N=13$ atom clusters  having one of these three 
crystal structures, since the absolute values of their cohesion energies 
are larger than those obtained for other  known simple structures (such as simple-cubic
or body-centered-cubic).

Since the  magnetism in RE metals comes from the localised 4f orbitals,  a  Heisenberg hamiltonian is a reasonable model. Different approaches exist to the treatement of the
 spin (classical or quantum approach) and to the interaction constants.  In~\cite{re4},   based on the fact that the calculated Fermi energy of the band structure is in a continuum of the density of states,
 a classical RKKY-Heisenberg hamiltonian is considered to describe the $Gd_{13}$ cluster.
 On the other hand in~\cite{lopez} both quantum and classical approaches of a RKKY-Heisenberg 
hamiltonian are studied for different closed-packed N=13 clusters and for different geometries
of the N=14 clusters.  Concerning the thermal magnetic behaviour of the N=13 
clusters, the results are qualitatively the same as in the classical approach developped 
in~\cite{re4}.

In this article we model the exchange interaction by a classical Heisenberg hamiltonian.
In addition, a surface anisotropy term that acts only on
the atoms at the surface to  account for the reduction
of the symmetry of the crystal at the surface is considered.

The main sources of anisotropy are the crystal-field anisotropy and the magnetic dipolar interaction.  The latter, being a long range interaction, can in principle be neglected for small clusters.
Choosing the correct term to model the surface anisotropy  of a RE cluster is not an easy task
as no experimental results are available. Nevertheless some hints can be taken from what is known for metallic clusters. 
In~\cite{luis} it has been found that the surface anisotropy term increases with decreasing
cluster size. 

When trying to  apply these results to RE clusters one can  notice that 
first principle calculations, which concern $Gd_{13}$,~\cite{hendriksen,re4} show
 a contraction of the surface layers that may induce a much larger crystal field anisotropy
at the surface than in the bulk. As it has been discussed in previous 
articles~\cite{hendriksen,kodama,aharony},
the lowering of the symmetry at the surface due
to the missing neighbours produces a 
crystal field with a predominant axial term in the radial direction which can 
be modelled by adding to the hamiltonian a 
term of the form $K_s S_\xi^2$ where $\xi$ is the component along the radial vector.

The classical RKKY-Heisenberg hamiltonian which is known to be a good model for 
bulk Gd~\cite{re4} has been already studied numerically in a simplified version 
(ferromagnetic  first-neighbour
interactions and antiferromagnetic interactions among {\it all} other pairs)  without 
anisotropy~\cite{cerovski}.

 In this article we consider a complementary approach. Our aim is to study the influence of the 
surface anisotropy energy. For simplicity, as a first approach, we  limit our model 
to the case of ferromagnetic first-neighbour couplings,
the same at the surface and in the bulk,
in order to avoid mixing the  effects of  surface anisotropy with those of competing interactions. It
can be argued that for a very small particle, the spins on  opposite faces of the surface are 
close enough  to interact with each other, thus modifiying the form of the anisotropy. As 
in the case of the second neighbours interactions, this effect has been neglected.
 Since we consider clusters having an approximately
spherical shape, we  have also neglected the shape anisotropy.

These considerations lead to  the following  hamiltonian which we study for the 
three  crystal lattices  (hcp, fcc, ico):

\begin{equation} \label{hamil}{\mathcal H}=-J\sum_{<i,j>}\vec{s}_i.\vec{s}_j-K_s \sum_{N_s}(\vec{s}_i.\vec{n}_i)^2
-\sum_{i=1}^N\vec{h}.\vec{s}_i,
\end{equation}  

where $\vec{s}_i$ is a classical Heisenberg spin with unit length,  $\langle i,j \rangle$
denotes the sum over all nearest-neighbour pairs, 
$N_s$ is the number of spins at the surface of the cluster (here  $N_s=12$), 
$\vec{n}_i$ is the unit vector giving the radial direction from the 
central spin, $s_1$.   $J>0$, $K_s>0$ and $h$ are the ferromagnetic interaction constant, 
the radial surface anisotropy constant, and the intensity of the applied magnetic field, measured 
in units of $k_B$ respectively. 
In this article we fix $J=1$.

\section{Calculation details}
 \label{sec:2}

The cluster is first simulated using 
 standard Monte Carlo simulations for the temperature magnetic dependence of the system
for the fcc, hcp and icosahedral lattices in zero field (heating and cooling simulations).

\begin{figure}
\begin{center}
\includegraphics[width=4cm]{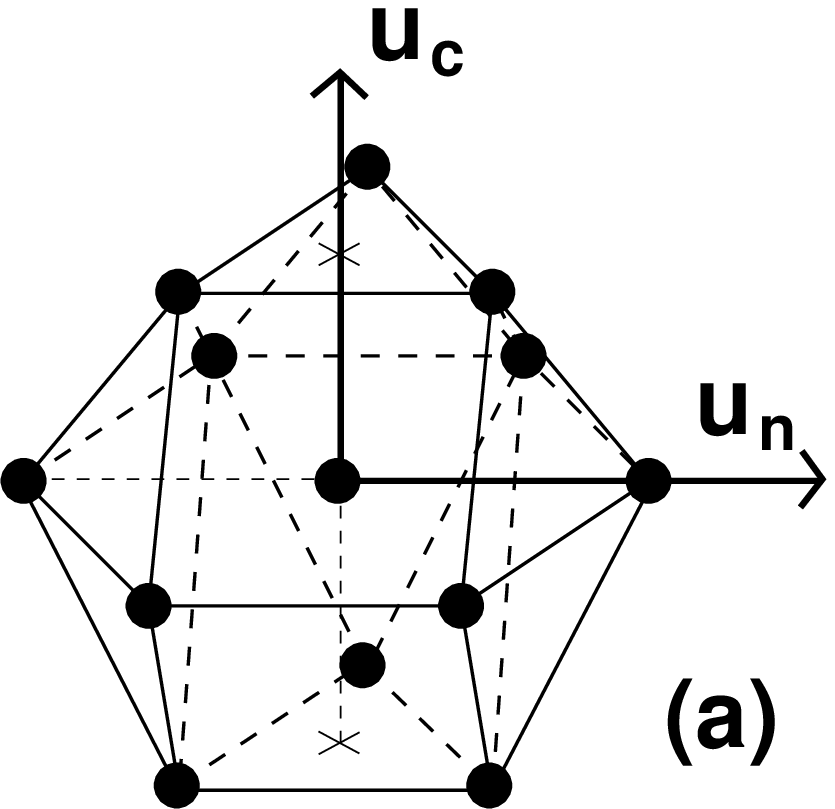}
\hspace{0.5cm} \includegraphics[width=4cm]{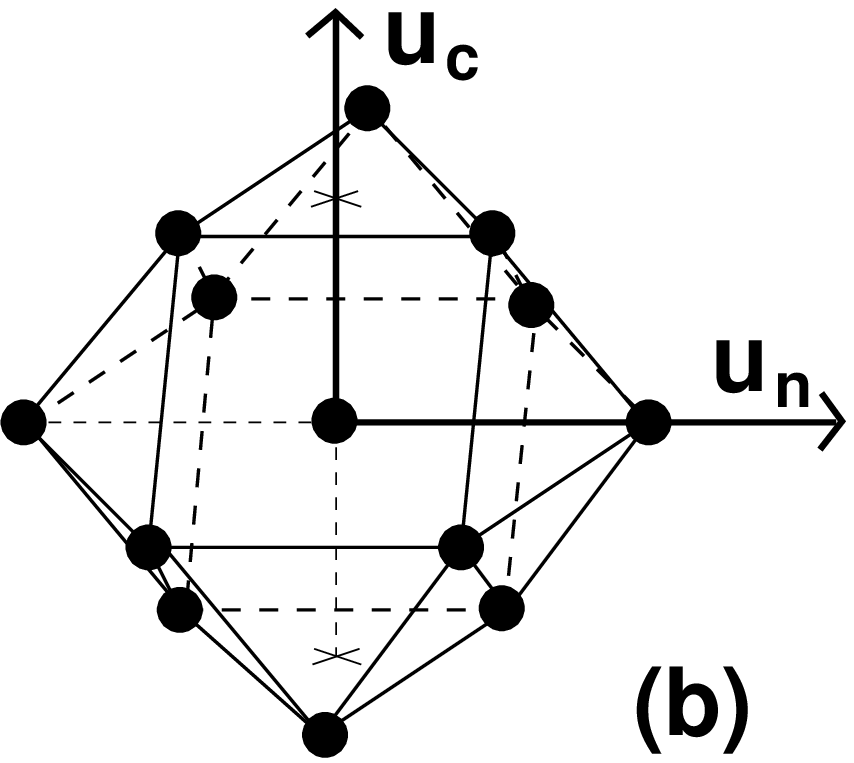}
\hspace{0.5cm} \includegraphics[width=4cm]{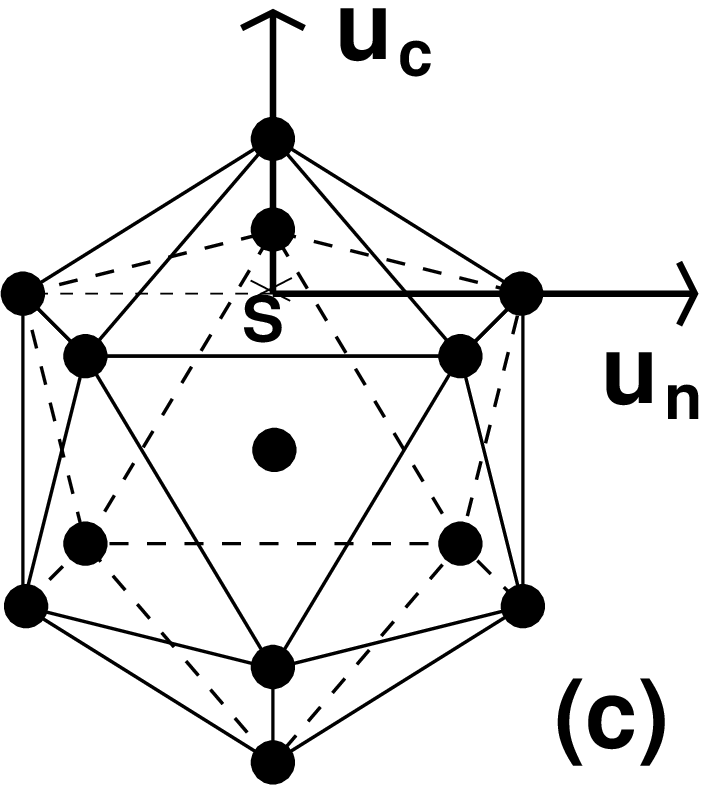} 

\caption{Studied lattice structures, definition of $\vec u_c$ and $\vec u_n$ directions of the
applied field for the field cooling and hysteresis simulations. (a) hcp, (b)fcc, (c) icosahedral lattices.}
\label{lattices}
\end{center}
\end{figure}

Heating and cooling cycles of the system in constant field are simulated.  
The different considered orientations of the applied field are depicted in figure
~\ref{lattices}.
For hcp lattice these orientations are either  parallel or perpendicular to the $\vec c$ axis
of the crystal.
The former is noted $\vec H_c=h \vec u_c$ and the latter
$\vec H_n=h \vec u_n$, where  $\vec u_n$ is a basis vector of the hexagonal central layer.
As the fcc lattice can be seen in a layered way,  the same definition of the 
direction of the applied  field as in the hcp structure is used
(see figure~\ref{lattices}.b).
For the icosahedral structure, the  $\vec u_c$ orientation is chosen along
the direction of one of the 5-fold symmetry axes. The complementary $\vec u_n$ direction 
is perpendicular to  $\vec u_c$ and joins the intersection point, noted S,  
of the 5-fold axis with the perpendicular plane containing the 5 icosahedral sites,  
with one of these sites (see figure~\ref{lattices}.c).
Hysteresis cycles are simulated at different  temperatures, with the magnetic 
field applied in the predefined directions of each structure.

All these simulations  are done using $10^6$MCS/s steps for calculation after having discarded
$10^6$MCS/s steps for the thermalisation process.  Quantities are averaged using 
one configuration every 100MCS/s in order to diminish statistical correlations.

We  define  the average cluster magnetisation as
\begin{equation}
\label{mag} m=1/N<| \sum_{i=1}^N \vec s_i|>,
\end{equation}

and the  average surface magnetisation of the cluster,
\begin{equation}
\label{smag}m_s=1/N_s< \sum_{N_s} | \vec s_i.  \vec n_i|>,
\end{equation}   

The  energy is also calculated as a function of the temperature and the magnetic field.

\section{Results}
 \label{sec:3}

\subsection{\bf Charaterisation of the ground states }
 \label{ssec:gs}

In order to have  a better understanding of the effect of  surface anisotropy 
at finite temperature and in an applied field, it is necessary to have an idea of
the ground state of the model for the different lattice structures in zero field. 
Unfortunatelly,  the  analytical determination of the  ground state of this hamiltonian
is not straightforward.
The coupling between the spins and the crystal structure makes it impossible to
perform a reduction of the number of variables in the problem by a global rotation of 
the magnetic structure.  
Moreover, as this coupling is non-linear, one can not use the local field orientation 
search methods for the ground state. 
We search for the ground state numerically using simulated-annealing, with  a decreasing
power law for the temperature: $T_{i+1}=(T_i)^\alpha$, where "i" labels the cooling step and 
 $1.05<\alpha<1.15$.  

We obtain ground-state magnetisations $m(0)< 1$ for the three
considered lattices due to the surface anisotropy term, which favors a canted
low temperature configuration.
This non-collinear order is confirmed by the surface average magnetisation which saturates 
near $m_s(0) \approx 1$ for big enough values of $K_s$.

 \begin{figure}
 \begin{center}
 \includegraphics[width=8cm]{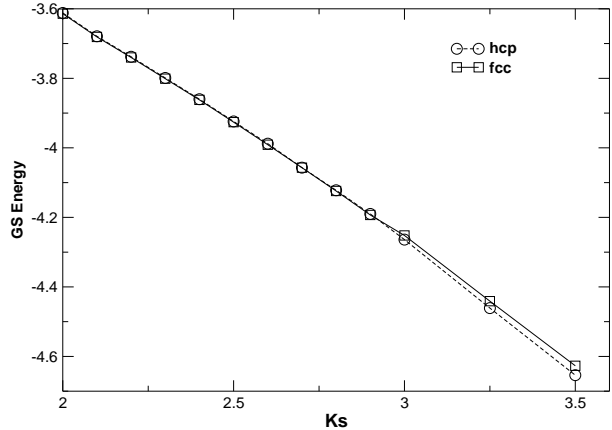}
  
  \caption{Ground-state energy as a function of $K_s$ from simulated annealing
  for the hcp and  fcc lattices.}
   \label{gs_k}
    \end{center}
     \end{figure}

We note a qualitative difference between the ground-state behaviour of the hcp and fcc clusters
and that of the icosahedral cluster.
The ground-state energies and magnetisations of the hcp and fcc clusters are very close 
for low anisotropy  and start to differ for a value around $K_s=3$,
while the corresponding energy and magnetisation values are very different for the 
icosahedral cluster.
Figure~\ref{gs_k} shows the ground-state energies for the two structures having the closest behaviour
in the interesting region of the anisotropy, $2<K_s<4$.  

\begin{figure}
\begin{center}
\includegraphics[width=4cm,clip]{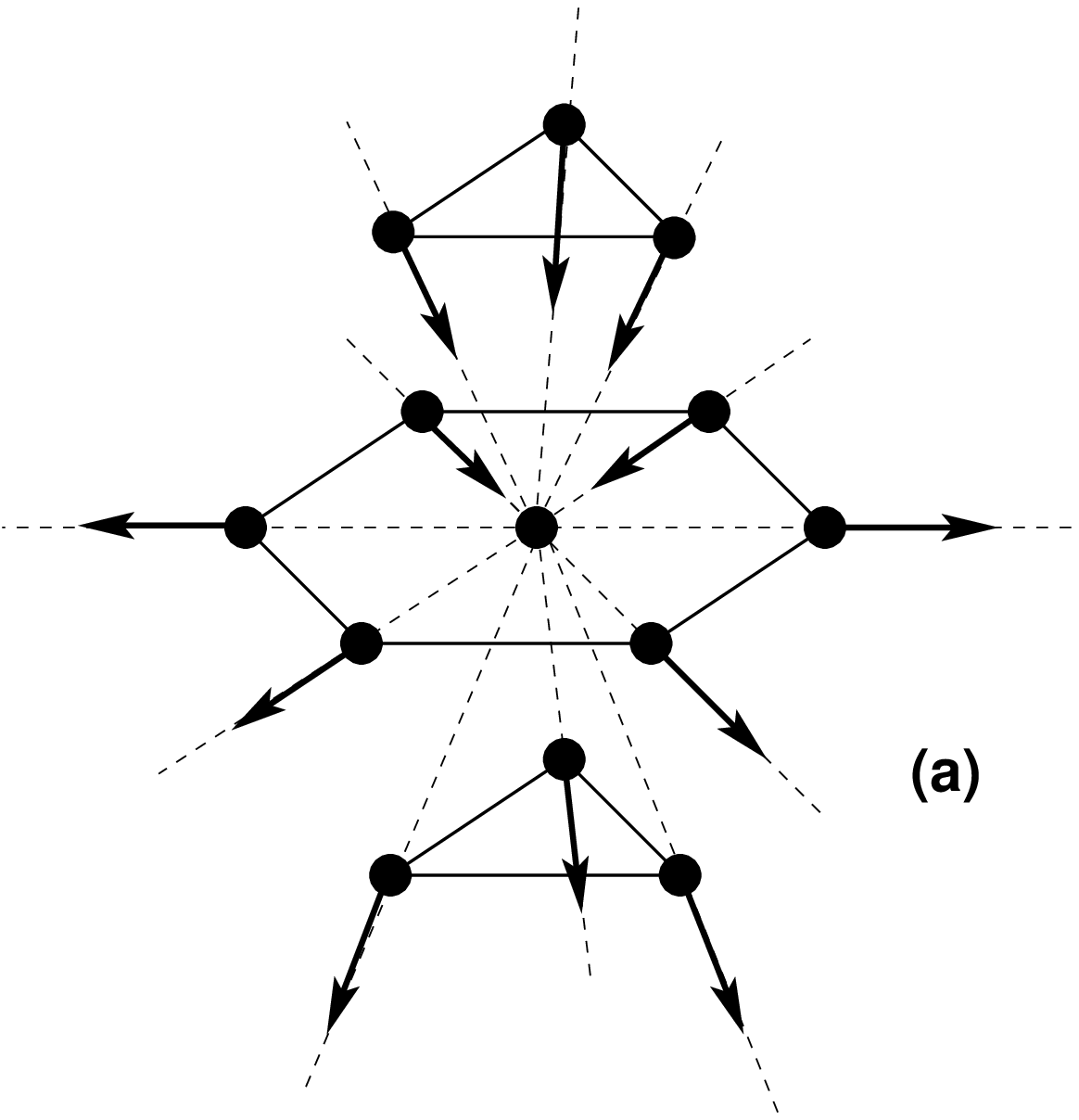}
\includegraphics[width=3.4cm,clip]{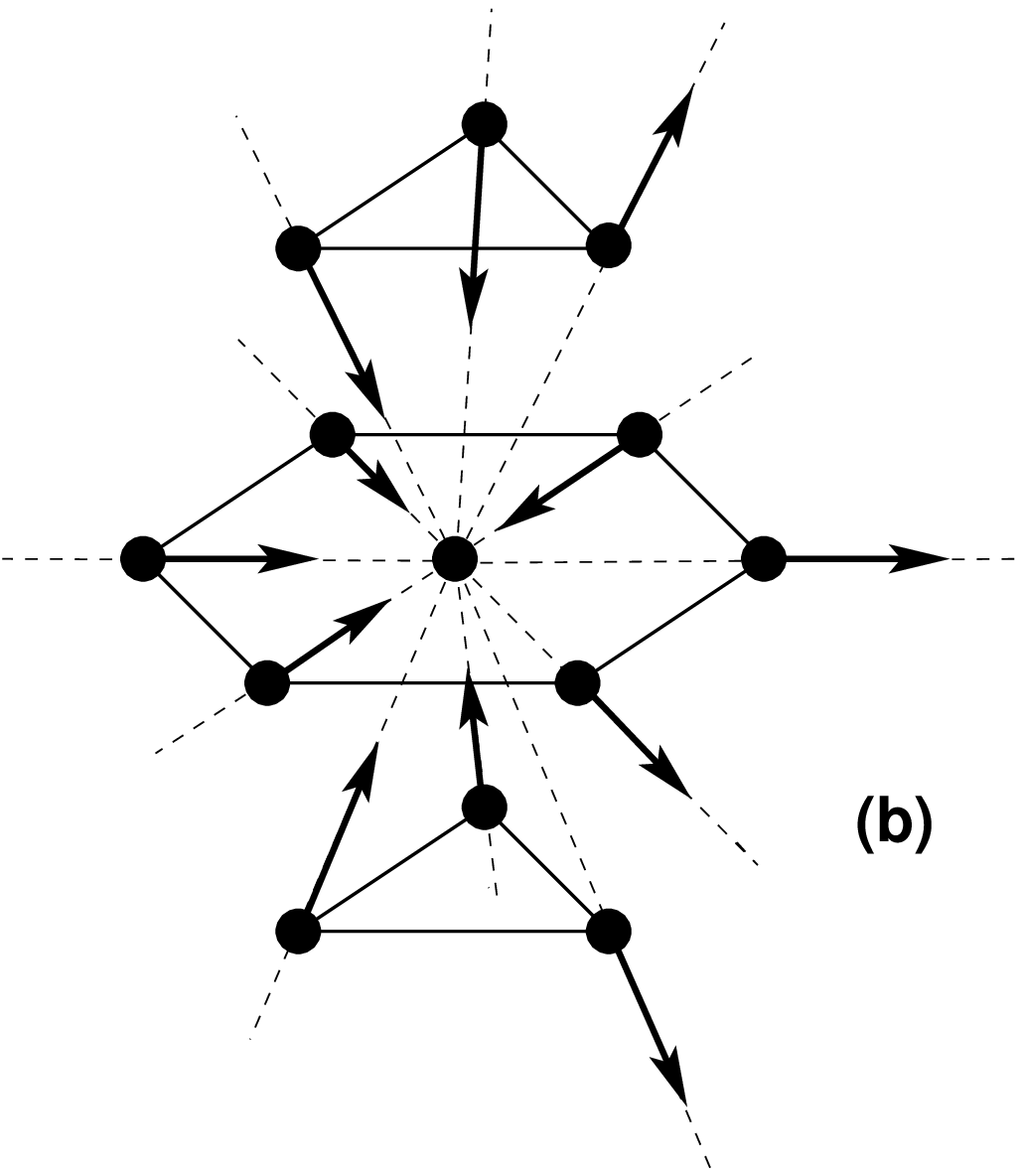}
 
 \vspace{1cm}
  
  \includegraphics[width=4cm]{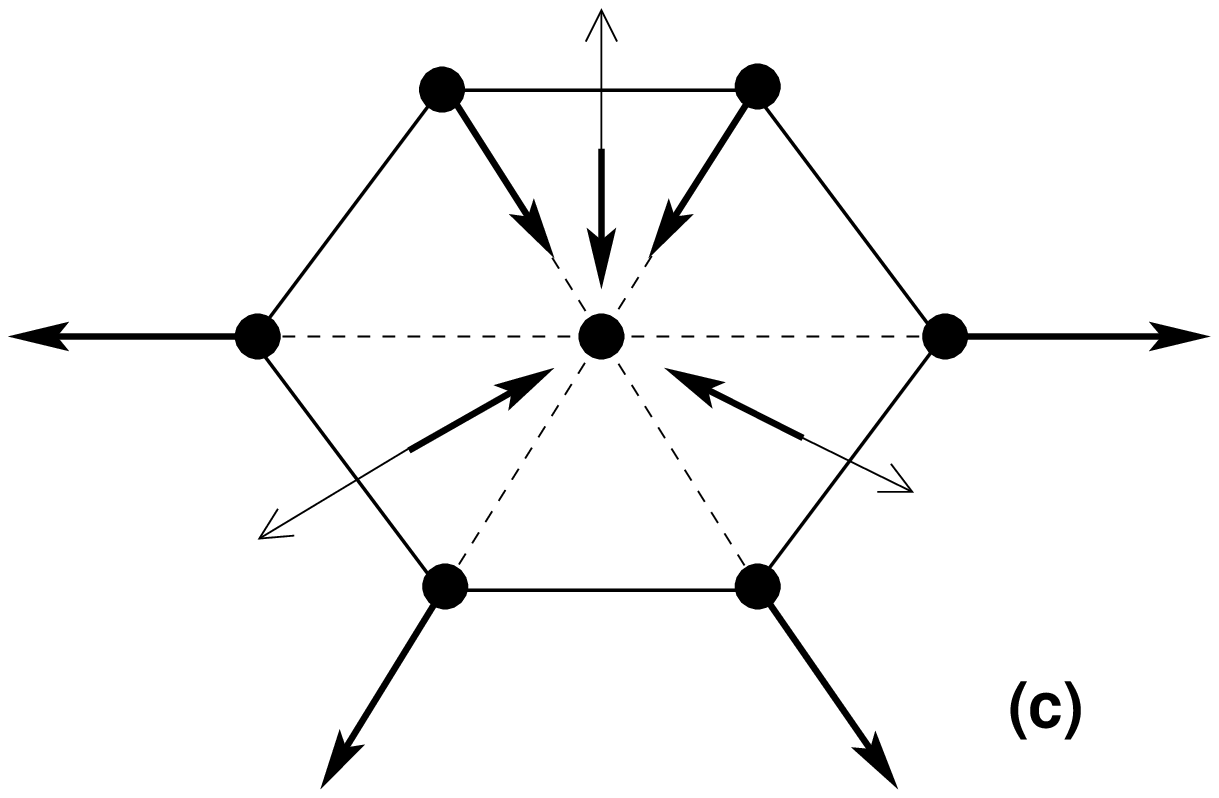}
  \includegraphics[width=3.4cm]{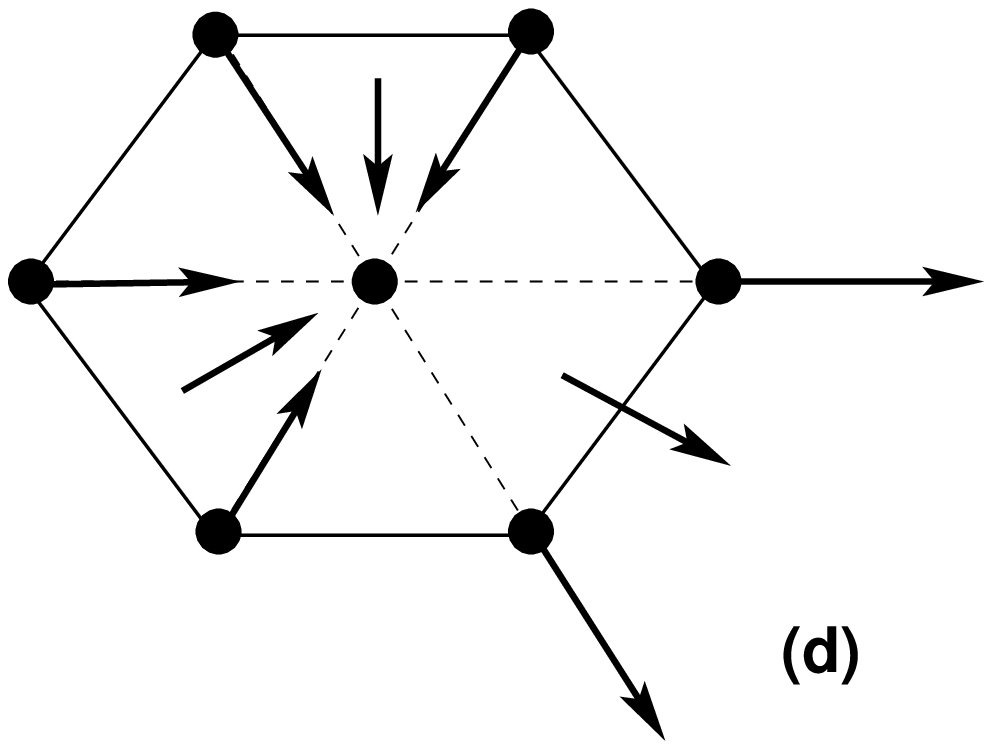}
   
   \caption{Typical ground-state configurations for the hcp lattice for $K_s=3$. Dotted lines indicate the
    radial directions. The real configurations show a slight deviation from the radial directions 
    which we haven't drawn here for simplicity. The central spin  is irrelevant (it follows
     orientation of the majority of its in-plane neighbours) and is not shown here for the sake of clarity.
    (a) Ground state corresponding to the case with the higher magnetic moment (mainly aligned along $\vec u_c$).
    (b) Ground state corresponding to the case with the lower magnetic moment (mainly aligned along $\vec u_n$).
    (c) Projection of configuration (a) on to the central plane; thick arrows indicate the upper plane 
     projections and thin arrows those of the lower plane.
     (d) Projection of configuration (b) on to the central plane; upper and lower planes projections 
     are coincident.}
     \label{gs_conf_hcp}
     \end{center}
     \end{figure}

\begin{figure}
\begin{center}
\includegraphics[width=4cm]{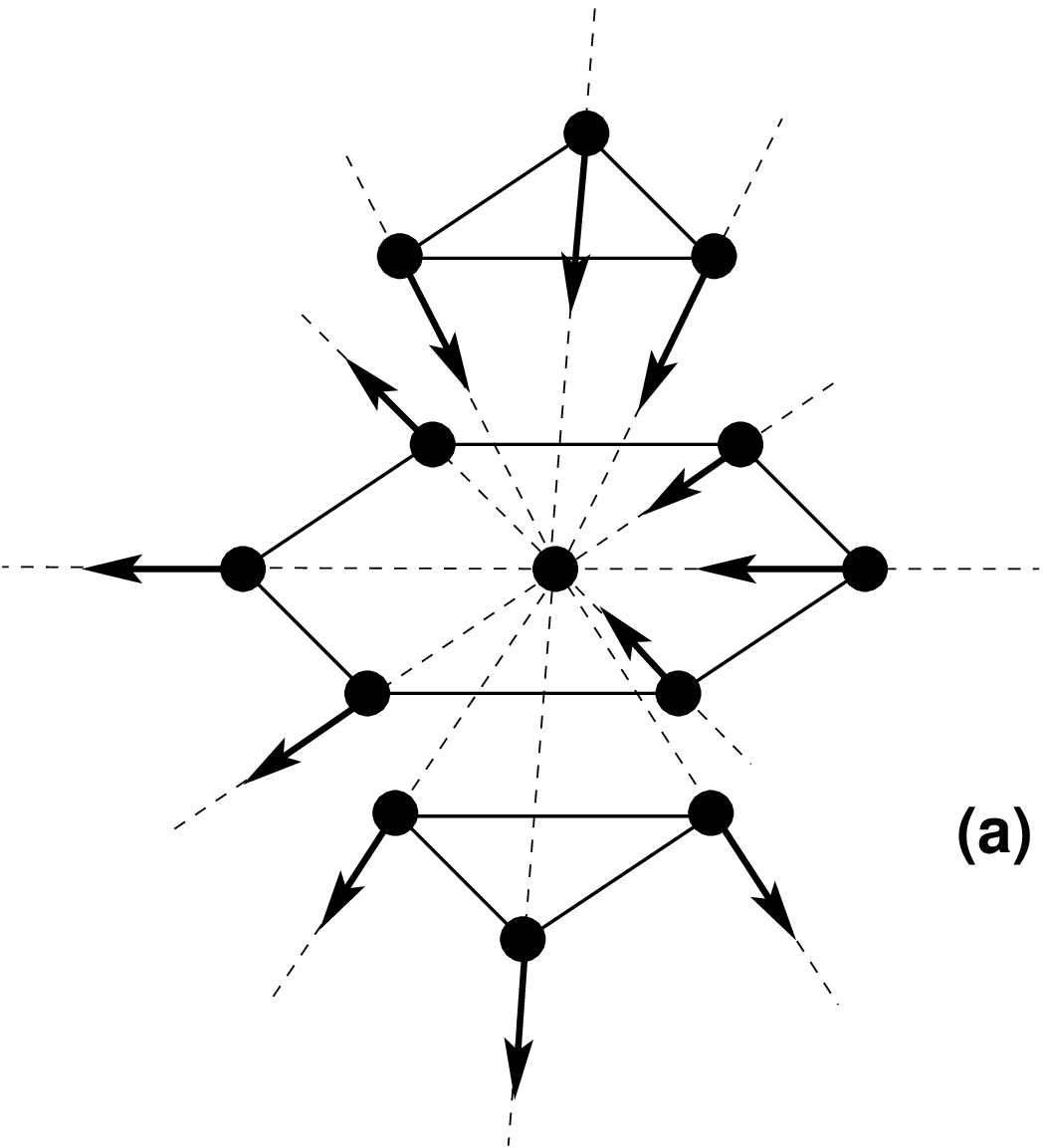}
\includegraphics[width=3.7cm,clip]{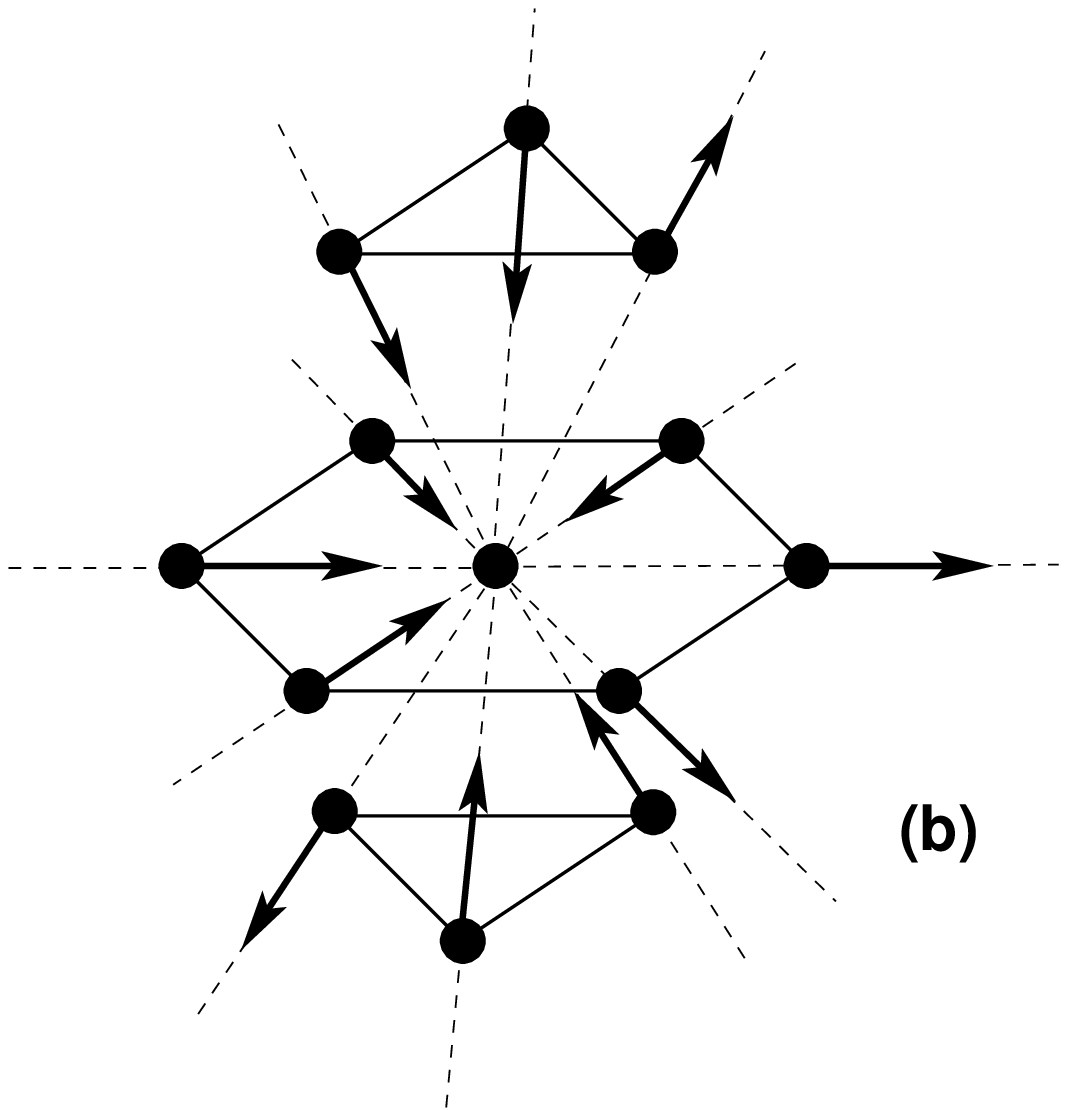}
 
 \vspace{1cm}
  
  \includegraphics[width=3.9cm,clip]{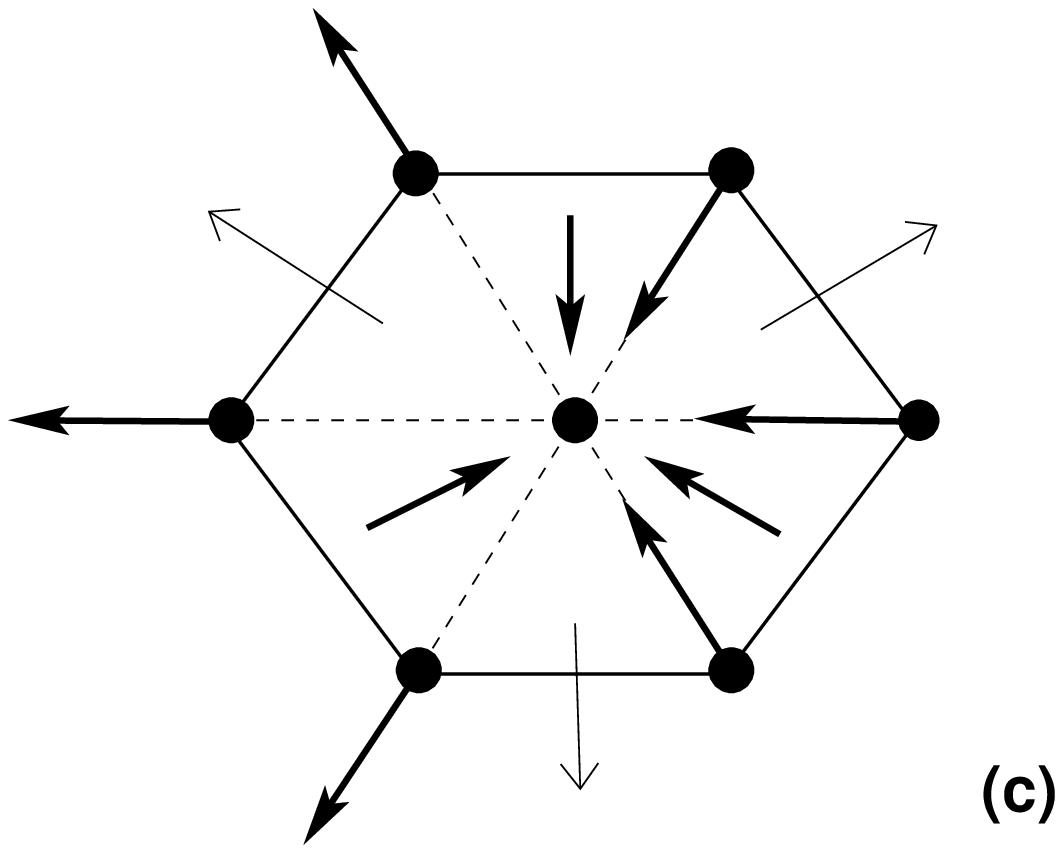}
  \includegraphics[width=3.9cm,clip]{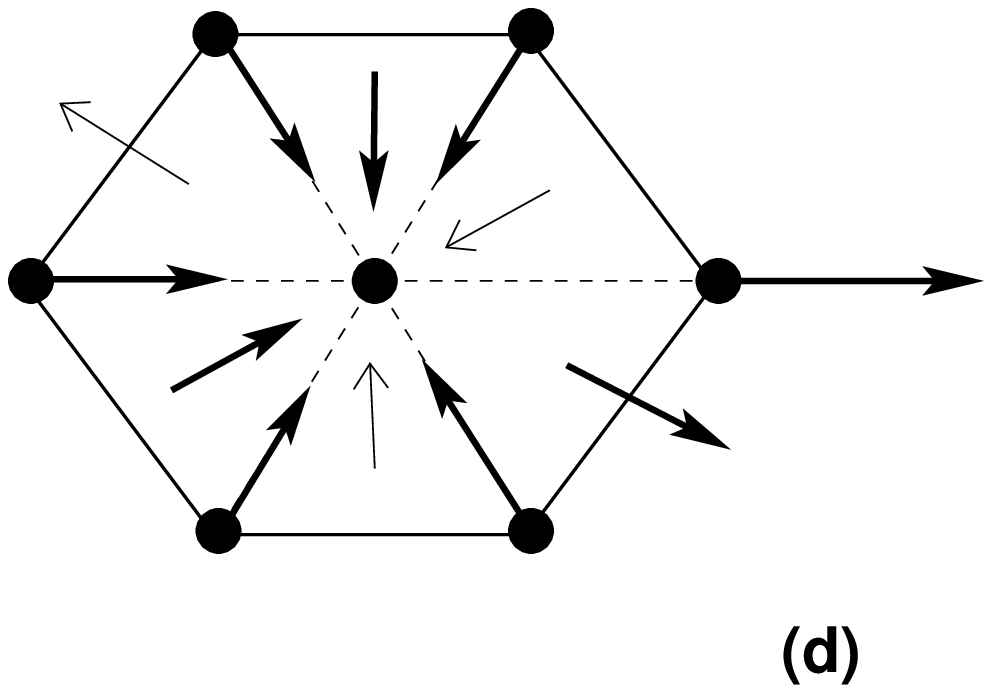}
   
   \caption{Typical ground-state configurations for the fcc lattice for $K_s=3$. Dotted lines indicate the
   radial directions. The real configurations show a slight deviation from the radial directions which
   we haven't drawn here for simplicity. The central spin is irrelevant (it follows
   orientation of the majority of its in-plane neighbours) and is not shown here for the sake of clarity
   (a) Ground-state configuration ; it has the same magnetic moment than the hcp ground-state 
    configuration (with the highest magnetic moment) given in figure~\ref{gs_conf_hcp}(a). 
   (b) ''Non-physical configuration'' obtained imposing the other hcp ground-state configuration
   (with the lowest magnetic moment) given in figure~\ref{gs_conf_hcp}(b) to the fcc lattice ; 
   the energy of such a configuration is much higher than the energy of (a).
   (c) Projection of configuration (a) on  to the central plane.
   (d) Projection of configuration (b) on  to the central plane ; thick arrows indicate the upper
   plane projections and thin arrows those of the lower plane.}
   \label{gs_conf_fcc}
   \end{center}
   \end{figure}

Figures~\ref{gs_conf_hcp} and~\ref{gs_conf_fcc} illustrate schematically the ground-state 
configurations found for the hcp and fcc structures for $K_s=3$, where the two lattices
start to differ.
For the hcp lattice we find two degenerate ground states corresponding to two different values of 
the magnetic moment.
For the fcc cluster, only one magnetisation state is found. 

Two typical low temperature configurations of the hcp lattice are represented in 
figures~\ref{gs_conf_hcp}(a) and ~\ref{gs_conf_hcp}(b).
In figures~\ref{gs_conf_hcp}(c) and ~\ref{gs_conf_hcp}(d) their projections on the central plane
may be seen (for  clarity, we draw the spin orientations along 
the radial directions;  the real spin configurations show slight deviations 
from these directions).
The magnetic moment of the cluster (shown in figure~\ref{gs_conf_hcp}(a)), 
corresponding to the ground state with the higher magnetisation,
is mainly aligned along $\vec u_c$, while for the case of the lower magnetisation
ground state (given in figure~\ref{gs_conf_hcp}(b)), the magnetic alignement is mainly along $\vec u_n$.

In figure~\ref{gs_conf_fcc}(a) we show a typical ground-state configuration of the fcc lattice. 
This configuration has the same magnetic moment  as the hcp ground state with the higher 
magnetisation (figure~\ref{gs_conf_hcp}(a)).
And as for this hcp ground state, the main direction of the magnetic moment is along $\vec u_c$. 

In order to  understand why the ground state is not degenerate for the fcc cluster while it is
degenerate for the hcp cluster, we impose the low temperature spin configuration 
found for the hcp lattice, shown in  figure~\ref{gs_conf_hcp}(b),  to the fcc lattice 
(respecting all the possible symmetries).
This ``non-physical'' configuration is shown figure~\ref{gs_conf_fcc}(b).
Comparing figures~\ref{gs_conf_hcp}(d) and ~\ref{gs_conf_fcc}(d),
it becomes clear why the ground-state configuration with the lower magnetisation in the hcp 
is not found in the fcc cluster: the coupling between the spins in the central  
plane and those in the lower plane  increases the energy in the case of the fcc structure.

As $K_s$ increases, the degeneracy of the hcp ground state is lifted in favour of the most canted 
ground state, i.e. with a lower magnetisation than the ground state  of the fcc cluster,
and the ground states of the fcc and hcp lattices separate in energy and in magnetisation.
The hcp cluster is better able to minimize both exchange and surface energy.

\begin{figure}
\begin{center}
\vspace{0.5cm}
\includegraphics[width=5cm]{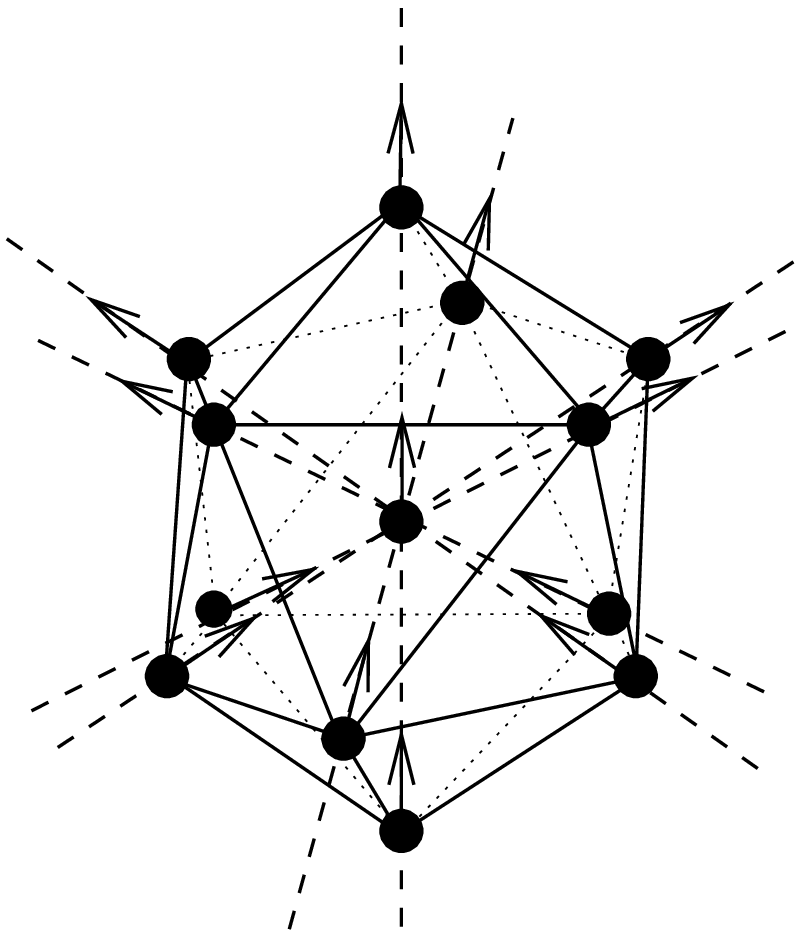}
\caption{Schematical ground-state configuration for the icosahedral lattice for $K_s=3$. Dashed lines indicate
radial directions. For the sake of clearness, we draw the spins fully oriented  along the radial
directions in spite of the fact that 
the actual configurations show a slight deviation from the  directions shown here.  This 
deviation tends to  align all the spins along the vertical $C_5$ axis, increasing the 
magnetisation with respect to the one of the pure radial configuration.}  

\label{gs_conf_ico}                                   

\end{center}
\end{figure}

The low temperature behaviour of the icosahedral cluster always differs from the other two structures.
The  magnetic moment values $m(0)$ of the icosahedral lattice at a given value 
of the anisotropy constant $K_s$ are higher than those of the other two structures 
in all the anisotropy range considered.
This can be understood by the different number of nearest neighbours at the surface. 
The sites at the surface of both hcp and fcc lattices have 5 nearest neighbours 
while the sites of the icosahedral lattice have 6.
At fixed $K_s$, the ferromagnetic coupling favours ground-state configurations with 
higher magnetisations in the icosahedral cluster than in the other two, in agreement 
with our results (see figure~\ref{gs_conf_ico}).
And the ground-state energies of the icosahedral lattice are much lower than the corresponding 
energies for hcp and fcc lattices.
This is due to the globular geometry of the icosahedral cluster that allow a better compromise 
to minimize both exchange and surface energy.

We will see that the similarities of both hcp and fcc clusters, as well as the specificities 
of the icosahedral cluster that we pointed out in this section, will be emphasized by the 
action of temperature or the application of an external magnetic field.

\subsection{\bf Zero field thermal behaviour}

\begin{figure}
\begin{center}
\vspace{0.5cm}
\includegraphics[width=10cm,clip]{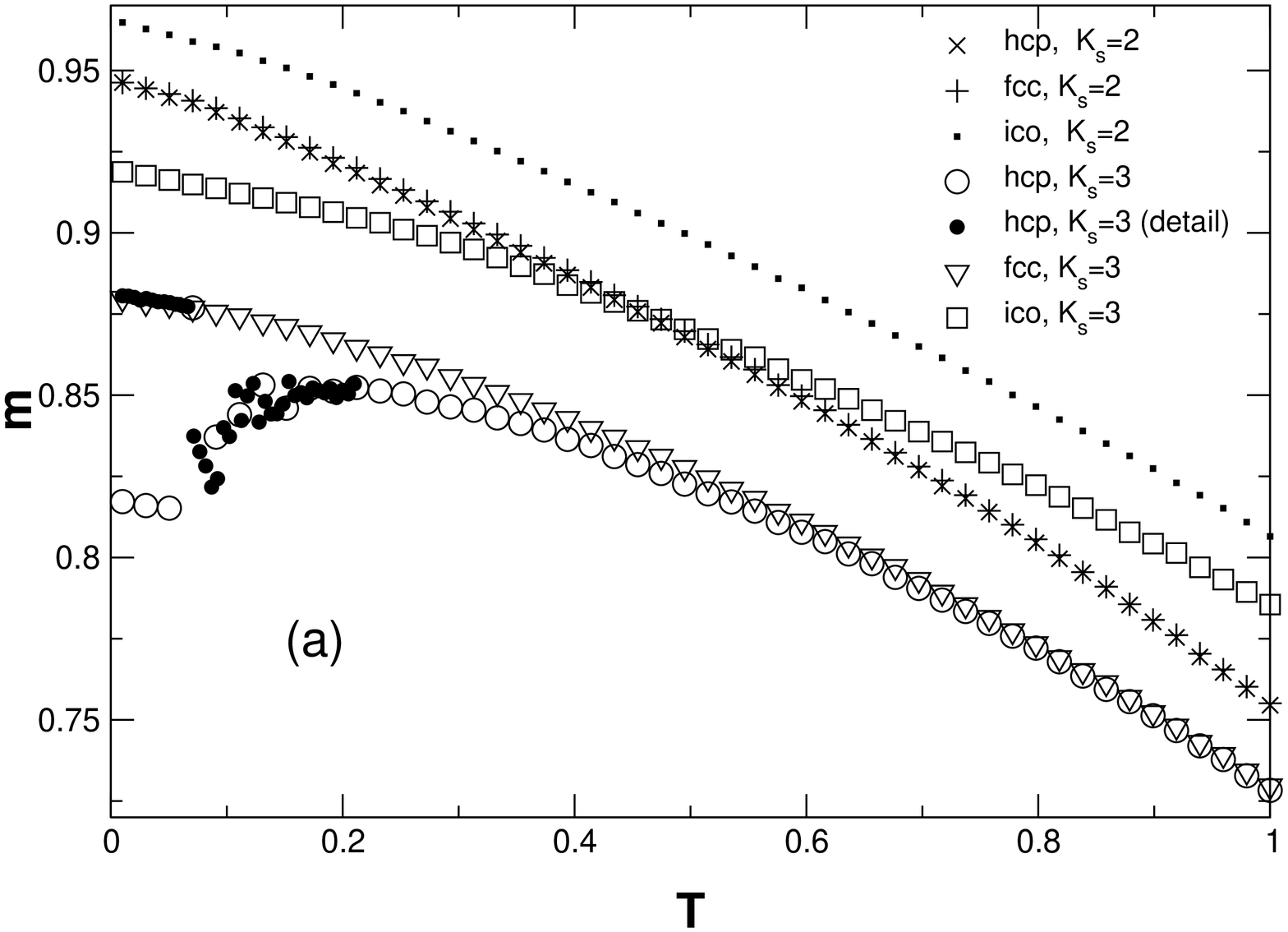}
 
\vspace{1cm}
 
\includegraphics[width=10cm]{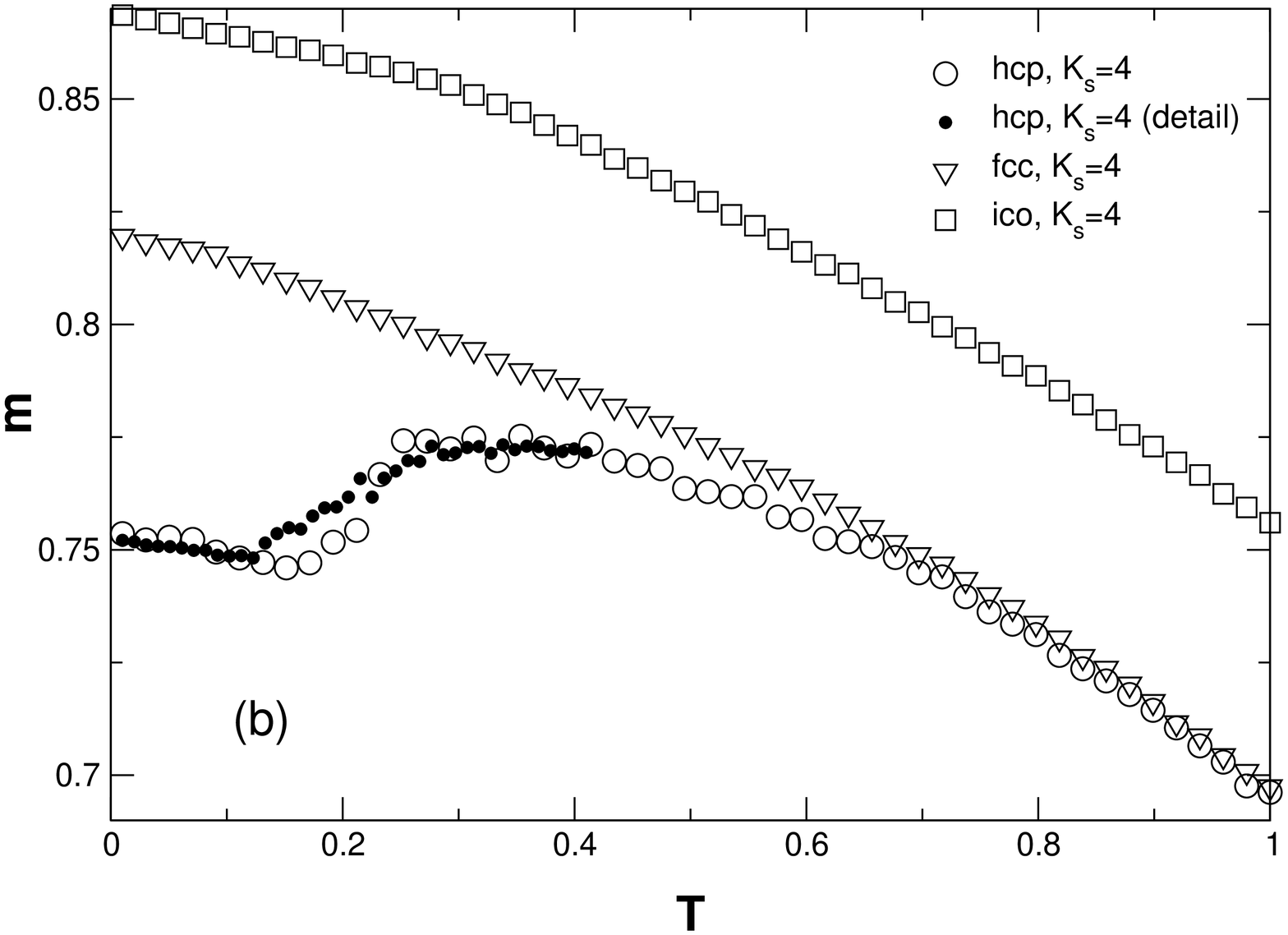}
\caption{Magnetic moment as a function of temperature for hcp, fcc and icosahedral lattice structure.
  Zero-field cooling simulations. Full symbols indicate a slow cooling.
  (a)  $K_s$=2 and 3.
  In a slow field cooling process for the hcp lattice with $K_s$=3, the magnetisation decreases when lowering $T$
  and suddenly increases again to rejoin the  fcc curve for the same  $K_s$.
  The two low $T$ states shown have the same energy.
  (b)idem,  $K_s$=4, the slow cooling process doesn't modify the peak and the saturation value
  of m is  lower than for finite $T$.}
\label{mt_h0}                                   

\end{center}
\end{figure}

We have performed heating and cooling cycles for the three considered lattices structures
in a range of values of the anisotropy constant $1\le K_s \le 7$.
Figure~\ref{mt_h0} shows the average magnetisation of the particle in a cooling process 
in zero field  for different $K_s$ values  and the three considered lattice structures.

The influence of  surface ansitropy  on the temperature behaviour of the magnetisation clearly 
depends on the crystal structure.
For $K_s=2$  the $m(T)$ curves increase monotonically for the three structures. 
The $m(T)$ curve of the icosahedral lattice is higher than those of the two other 
structures in all the measured temperature range.

For $K_s=3$ the degenerate ground state of the hcp structure leads to a $m(T)$ curve showing an
anomalous behaviour. Which of the ground states is reached  depends on the trajectory performed 
in the phase space while cooling the system.
As the temperature is lowered, the magnetisation of the cluster grows, but at a value $T_a$
it rapidly falls down creating a peak (open circles in figure~\ref{mt_h0}). 
If the cooling process is performed very slowly, the cluster magnetisation first decreases and then
increases again reaching the same low temperature value of the magnetic moment than for the fcc 
structure (full circles in figure~\ref{mt_h0}). 
Energy curves $E(T)$ of the two runs (not shown here) superpose exactly, and superpose also to the $E(T)$ 
curves of the fcc cluster.
Obviously, for the fcc where no degeneracy has been found, the behaviour of the $m(T)$ curves 
doesn't show any dependence on the trajectory in the phase space during the zero-field cooling proccess.
As expected, the two $m(0)$ values of figure~\ref{mt_h0} for hcp and fcc lattices correspond to the  
magnetisation ground states found in figures~\ref{gs_conf_hcp} and~\ref{gs_conf_fcc}.

When performing a standard MC  zero-field cooling simulation of the hcp lattice starting from the low magnetisation 
ground states found by simulated annealing for $K_s=3$, the zero-field cooling curves are reproduced exactly 
and the peak shown in figure~\ref{mt_h0}(a) appears again.

As we have discussed in the previous section, as $K_s$ increases, the degeneracy of the hcp ground 
state is lost and the ground states of the fcc and hcp lattices separate in energy (see figure~\ref{gs_k}).  
For instance, for the hcp cluster and $K_s$=4, the  peak is always found when cooling the 
system and as expected, the saturation value of the cluster magnetisation is smaller than for 
the fcc structure (figure~\ref{mt_h0}(b)).
Increasing $K_s$ diminishes the height of the peak till it disappears. 
This behaviour is a particularity of the hcp structure.

For the other lattice structures the cluster magnetisation increases monotonically when cooling 
and only for big values of  $K_s$ a plateau is observed. 

In all the cases the $m(T)$ curves are ordered (from highest to lowest case) as follows: 
icosahedral, fcc and hcp.  
This fact shows that, at fixed  $K_s$, a different degree of competition appears between the exchange 
and the surface term according to the considered lattice structure. 
We will see this fact emphasized by the application of an external magnetic field.

\subsection{\bf Non-zero field thermal behaviour}

We simulated the field cooling of the system with the field applied in directions described 
in figure~\ref{lattices}.  We  studied various intensities of the field ranging from
 $0 \le h \le 1$.

\begin{figure}
\begin{center}
\includegraphics[width=10cm,clip]{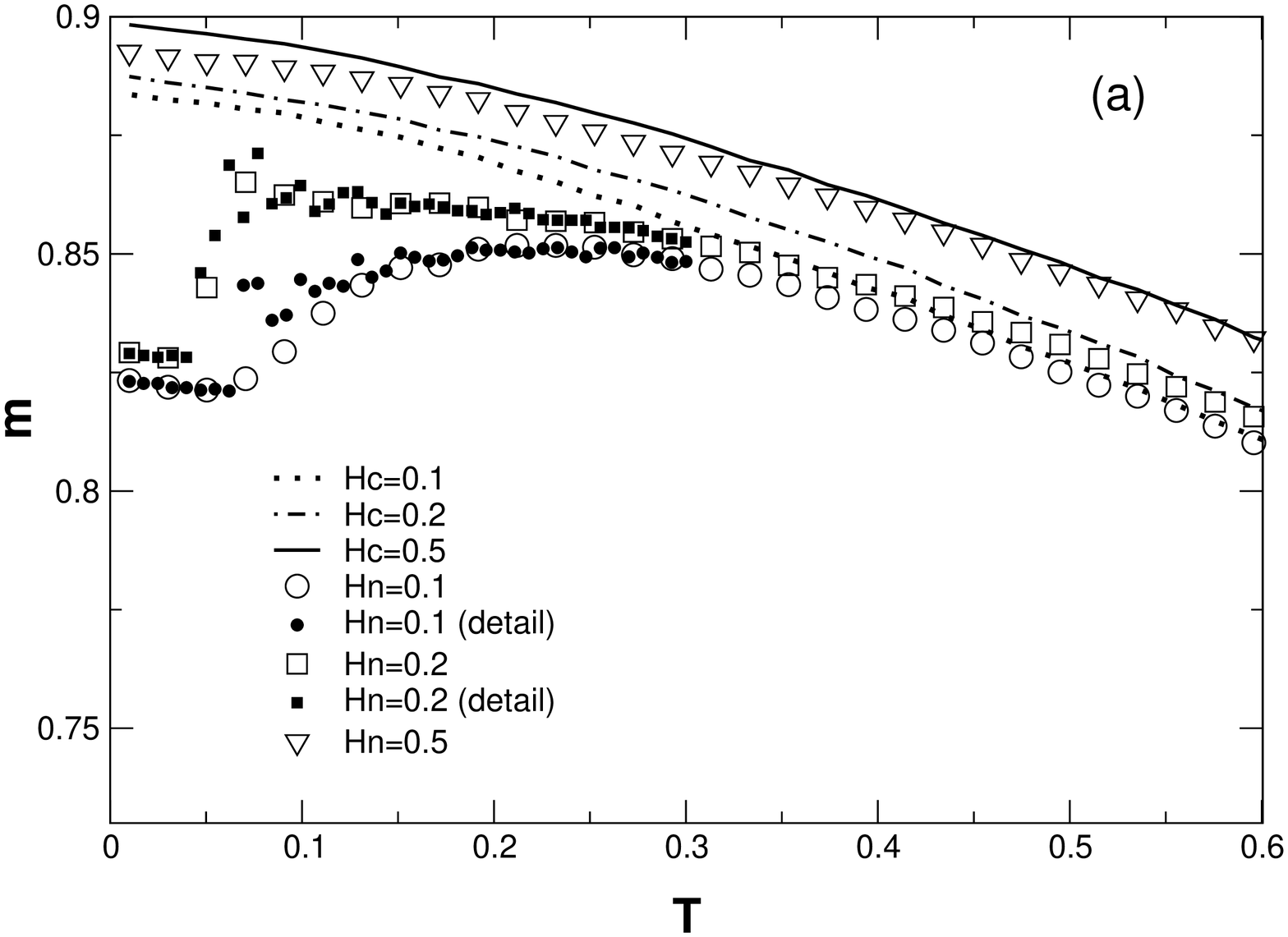}
 
\vspace{1cm}
  
\includegraphics[width=10cm]{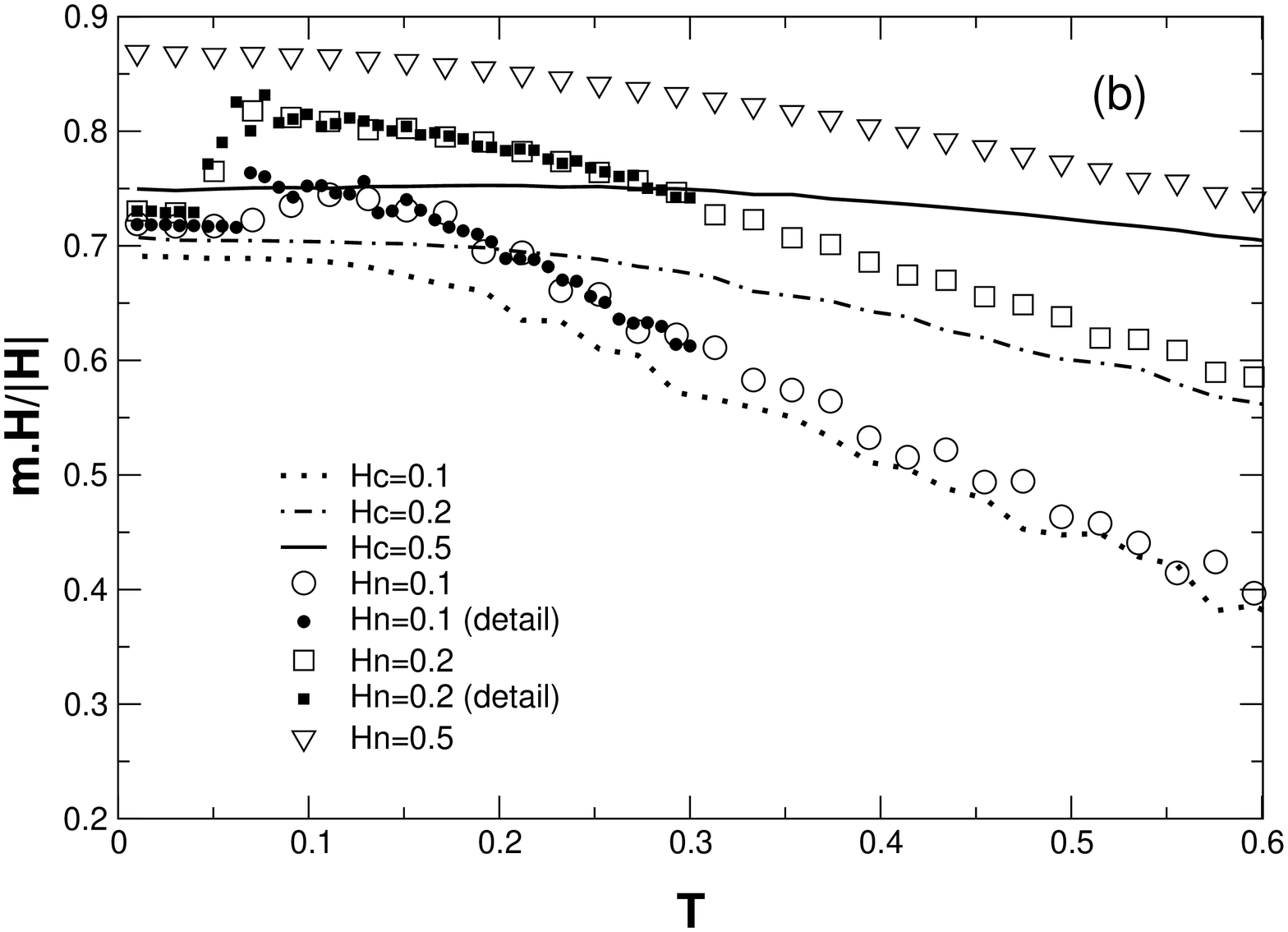}
\caption{Hcp lattice,  $K_s$=3. Field-cooling simulations. $H_c$ and $H_n$ denote the 
  magnetic field intensities along the directions described in figure~\ref{lattices}(a). 
  Full symbols curves, issued 
  from slow field-cooling simulations, show that the peak is stable. 
  (a)  Magnetic moment as a function of temperature. (b) Projection of the magnetisation along the
  axis of the corresponding applied field.}
\label{mT_hcp}
\end{center}
\end{figure}

Figure~\ref{mT_hcp}(a) shows $m(T)$ for  $K_s$=3 and the two
considered orientations of $\vec H$. When the field is 
$\vec H_c=h \vec u_c$   ($h \ge 0.05$) the magnetisation grows 
monotonically as the temperature is decreased, and the anomaly 
observed in zero field disappears. On the contrary, 
when the field is pointing in the $ \vec u_n$ direction, a peak is observed for the same $T \approx T_a$ 
where the anomaly is found in  zero field. 
This peak is confirmed by a very slow cooling process as reported in figure~\ref{mT_hcp}(a). 
Obviously, when the applied field is too high the peak is destroyed, and the magnetisation 
grows monotonically as $T$ is lowered.

The ground-state analysis of section~\ref{ssec:gs} helps to understand why this is so.  As decribed above,
two degenerate states are found for the hcp lattice for $K_s=3$ and $h=0$:  one corresponding to the monotonous $m(T)$
behaviour (figure~\ref{gs_conf_hcp}(a)) and another one corresponding to the peak in $m(T)$ curves 
(figure~\ref{gs_conf_hcp}(b)).  The former has its magnetic 
moment mainly oriented along $ \vec u_c$ and the latter mainly along  $ \vec u_n$.  
Then when the magnetic field is applied, one of these degenerate zero-field ground states 
is selected according to the orientation of the field.
For the fcc lattice, on the other hand, only one possibility exists (figure~\ref{gs_conf_fcc}(a))
and the $m(T)$ curves do not depend on the field direction.

In figure~\ref{mT_hcp}(b) we plot  $|\vec m .\vec H|/|\vec H|$ as a function of $T$ 
for different intensities and
the two studied orientations of the applied field. 
It can be seen that the behaviour of the corresponding projection of the average
magnetisation along the field direction depends on this direction.
When the field is applied along  $\vec u_n$ the corresponding projection of the average
magnetisation, called  $m_n$, easily follows the field as $T$ is lowered. However 
when the field direction is $\vec u_c$ the corresponding projection,  $m_c$, grows very slowly 
as $T$ is lowered saturating for $T<T_a$.

Comparing figure~\ref{mT_hcp}(a) and (b) one can understand the difference in the 
magnetic behaviour between  the two chosen field directions. 
When the field is in the $\vec u_c$ direction,  the magnetic moment grows monotonically (the peak
observed in zero field disappears) while the $m_c$ projection saturates at  value lower than one
at finite $T$.
This means that there is a non zero contribution to the magnetic moment  which is not in the 
$\vec u_c$ direction. 
The situation is different when the field is in the $\vec u_n$ direction, we can see that it is mainly $m_n$ 
which is responsible for the peak on the $m(T)$ curve.

\begin{figure}
\begin{center}
\includegraphics[width=10cm,clip]{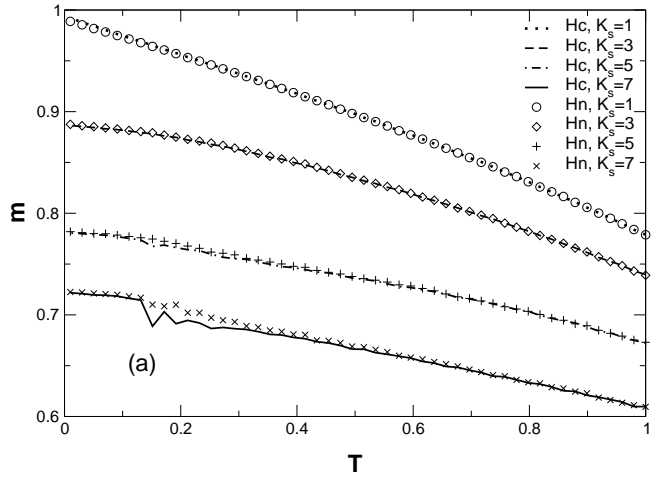}
 
 \vspace{1cm}
  
  \includegraphics[width=10cm]{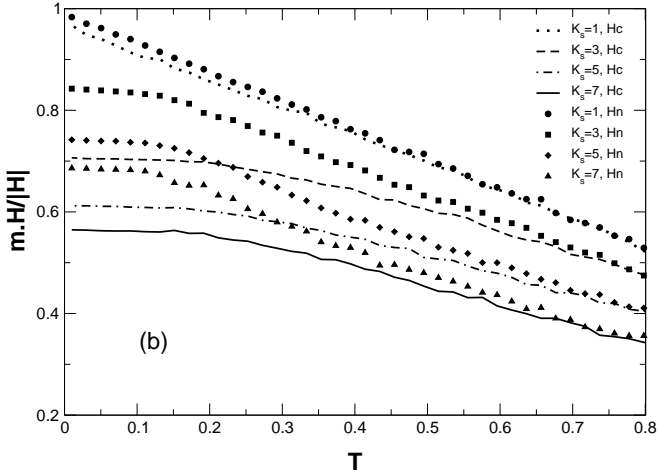}
  \caption{Fcc lattice, the intensity of the applied field is $h=0.2$.  Field cooling
  simulations.
  $H_c$ and $H_n$ denote the applied magnetic field
 parallel to the $\vec u_c$ and to the  $\vec u_n$ axis respectively
 (see figure~\ref{lattices}(b)).
    (a)The magnetic moment as a function of temperature is not affected by the direction of the
    applied field for all  $K_s$ values.
    (b) Projection of the magnetisation along the direction of the corresponding applied field. For all  $K_s$ 
    values,
    the magnetisation follows the field more easily when it is applied along  the $\vec u_n$
    axis.}
    \label{mT_fcc}
    \end{center}
    \end{figure}

For the fcc structure the behaviour is  different. To compare with the hcp lattice we 
considered the field directions as shown on the figure~\ref{lattices}(b). 
The $m(T)$ curves are coincident 
for both directions of the applied field and no anomaly is observed (see figure~\ref{mT_fcc}(a)).  
This result remains true for all field intensities  and for all values of  $K_s$.  
This can be understood in terms of  the structure of the ground state: for the  fcc lattice
there's no degeneracy of the ground state so the  $m(T)$ curves do not depend on the field direction.
On the other hand, as for the hcp lattice, when the field is applied in the $\vec u_n$ direction, 
$m_n$ follows the field  more easily than does $m_c$ when the field is applied
in the $\vec u_c$ direction (see figure~\ref{mT_fcc}(b)). 
The zero-temperature value of $m_c$ is lower than the corresponding one 
for $m_n$ for all  $K_s$, also in agreement with the hcp case. This can also be related to the 
characteristics of the ground-state configurations in zero-field (figures~\ref{gs_conf_hcp} and~\ref{gs_conf_fcc}):  
in both hcp and fcc ground
states  we have found that the spins in the central plane have a small component along the 
$\vec u_c$ direction, so, when a magnetic field is applied along  $\vec u_c$, the spins in the central plane find it 
more difficult to align with the field direction.

\begin{figure}
\begin{center}
\vspace{0.5cm}
\includegraphics[width=10cm,clip]{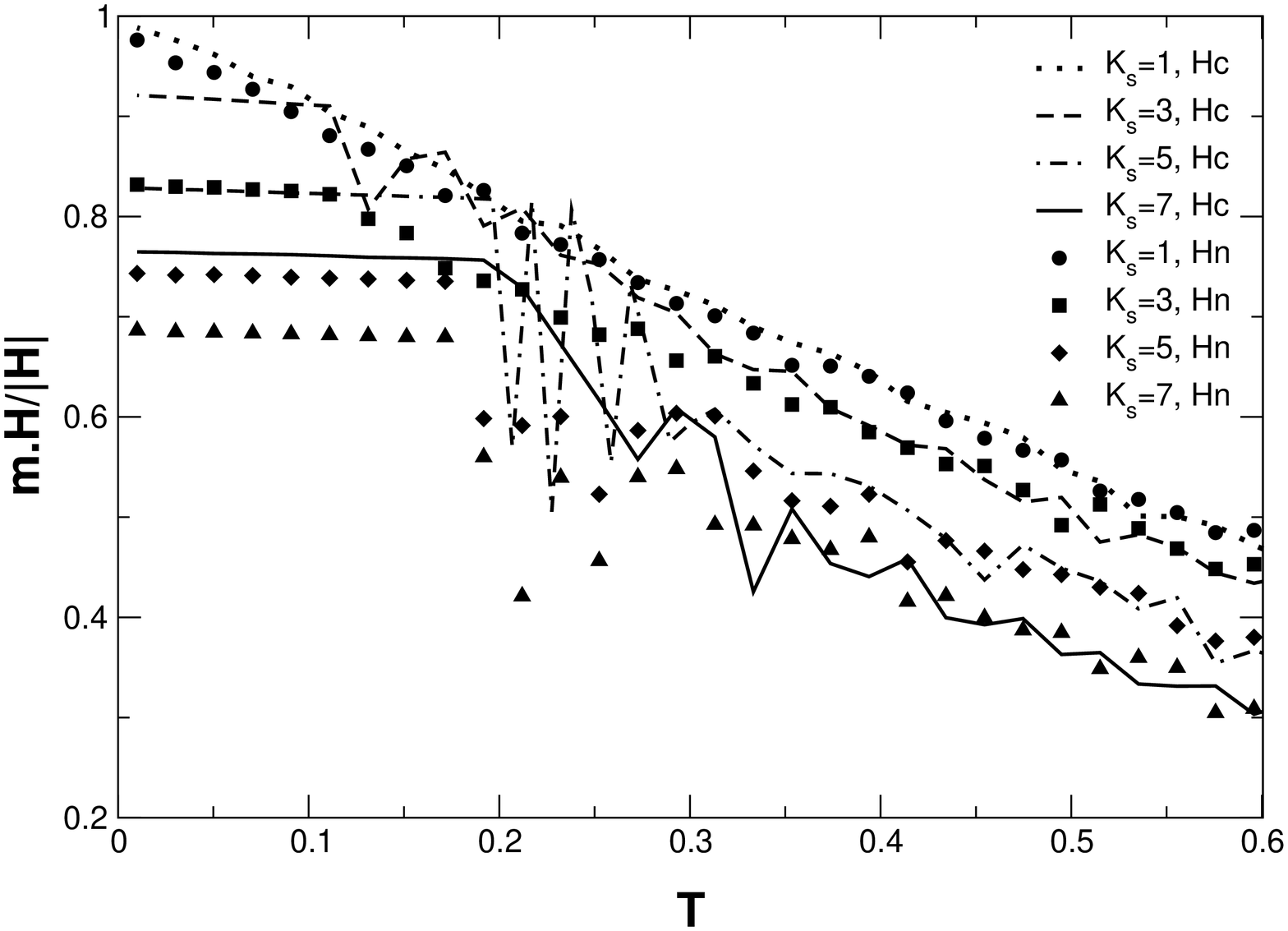}
 
\caption{Icosahedral lattice, the intensity of the applied field is $h=0.1$.  Field cooling simulations.  
$H_c$ and $H_n$ denote the magnetic field
applied parallel to the $\vec u_c$ and to the  $\vec u_n$ axis respectively
(see figure~\ref{lattices}(c)).
A plateau appears at  $T=T_s$ ($T_s$ increases with $K_s$).  For high values of  $K_s$   
( $K_s>5$) large fluctuations are observed just before the plateau.}
\label{mT_ico}
\end{center}
\end{figure}

The situation for the icosahedral lattice is less straightforward.  First,  this 
lattice has a globular structure rather than the layered structure of hcp or fcc, and so, one
cannot directly identify the 5-fold symmetry with the $\vec u_c$ axis of the other two layered 
structures.  Nevertheless, the results of  field cooled simulations show, as for the fcc case, 
that the orientation of the field has no influence in the $m(T)$ curves. On other hand, the 
orientation of the field affects the projection of the magnetisation in the field direction 
(see figure~\ref{mT_ico}). 
For this structure, it is $m_c$ which follows more easily the applied field. For $K_s \ge 3$  $m_c$
reaches its saturation value at a finite value of  $T=T_s$. This value increases with  $K_s$.
This is also a sign of a non-collinear state, confirmed by the high values of
the surface magnetisation at zero temperature ($m_s(0) \ge 0.9$ for $K_s \ge 6$).

For a given lattice and a fixed value of the field intensity $h$, two 
different regimes are observed for the thermal behaviour of  magnetic moment, $m(T)$, and 
of the projection of the magnetisation of the surface spins along the radial directions $m_s(T)$, 
  according to the value of  $K_{s}$.  For low  $K_{s}$,
$m(T)$ is  higher than  $m_s(T)$ for all $T$. For high  $K_{s}$ values,
the opposite 
behaviour is observed.  There is then a range of  $K_{s}^{c}$  values where the two curves 
cross at a 
finite temperature, $T_{cross}(h)$, leading to  a non-collinear ground state ($m_s(0)>m(0)$).  
In spite of the fact that one cannot actually talk of a phase transition,
this crossing indicates that the non-collinear state, where each magnetic moment has
a strong component perpendicular to the surface, takes over from a state of large global
magnetisation (collinear state). 
The range of $K_{s}^{c}$ values 
is  characteristic of each lattice structure. For the studied values of  $K_{s}$,
we have found  $K_{s}^c \approx 4$ 
for hcp and fcc and  $K_{s}^c \approx 5$ for the icosahedral structure. 
This  dependence of the  $K_{s}^c$ is related to   the different number of
nearest neighbours at the surface for the different structures.
As already mentioned, the sites at the surface of both hcp and
fcc lattices have 5 nearest neighbours while the sites of the icosahedral lattice
have 6, which leads to a 
larger  $K_{s}^c$ in this latter case so as to counterbalance
the ferromagnetic coupling.

We have studied the   $T_{cross}(h)$ dependence for a given lattice at the corresponding $K_{s}^c$ 
values, when  the field is applied along the $\vec u_c$ direction (highest symmetry axis for the
three lattices).  We observe that    $T_{cross}(h)$  shifts to low temperatures
as the applied field increases.  This simply  reveals the fact that when a field is applied, the 
thermal energy needed to flip from the  non-collinear to the collinear  state is lower.  In 
the case of very high fields,  $T_c(h) \to 0$, indicating that the ground state is 
already collinear.

\subsection{\bf Hysteresis loops.}

Starting from a zero-field cooling state, we performed hysteresis loops with the field oriented
along each of the directions described above.
We  observe that above the temperature  $T_a$ corresponding to the anomaly of the hcp lattice,  
no hysteresis is found. 
Then, to allow for comparison, we performed for all the lattices the hysteresis cycles at  $T \le 0.1$, 
where the hysteresis is found in hcp structure.

\begin{figure}
\begin{center}
\vspace{0.5cm}
\includegraphics[width=10cm,clip]{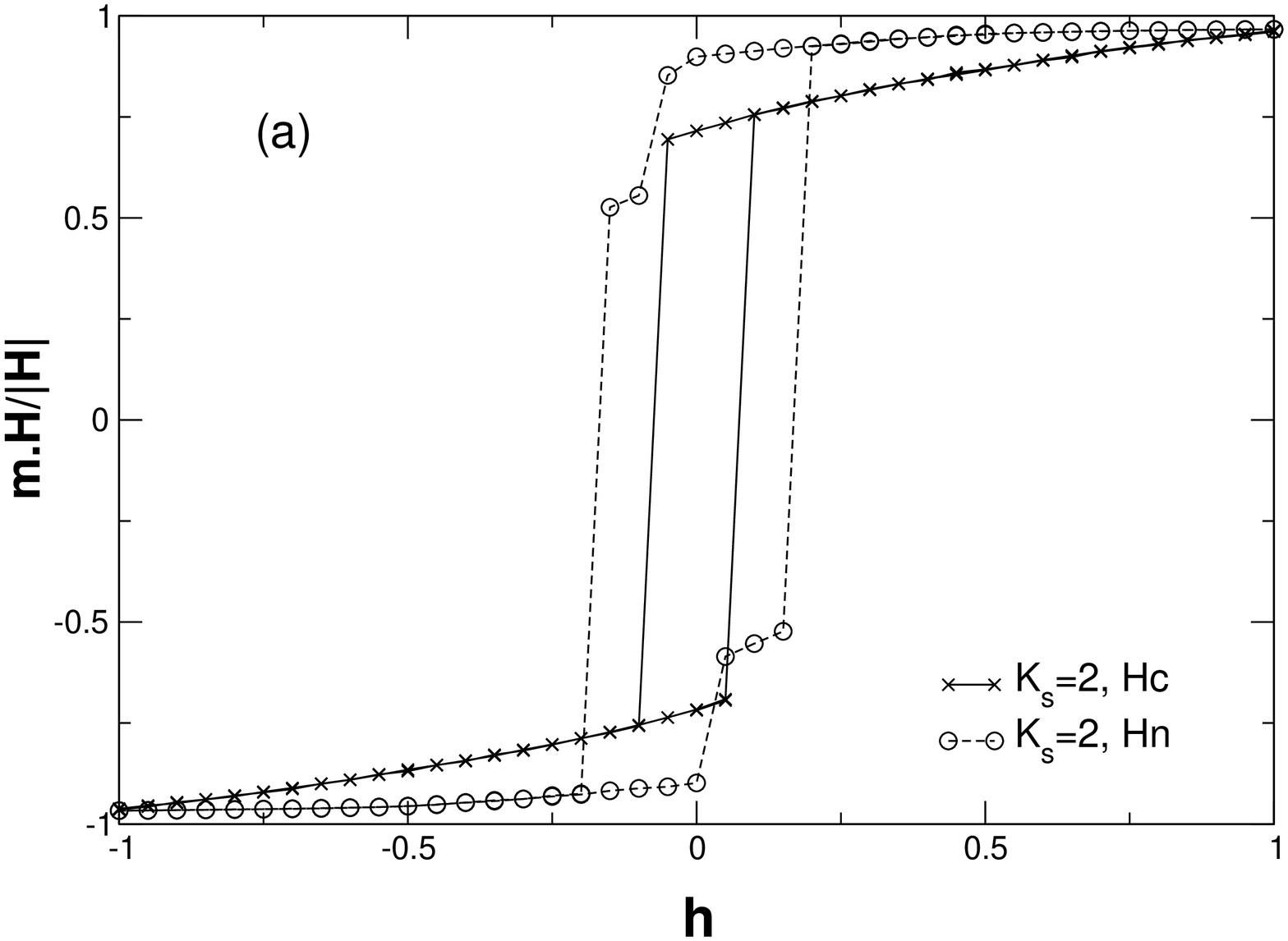}

 \includegraphics[width=10cm]{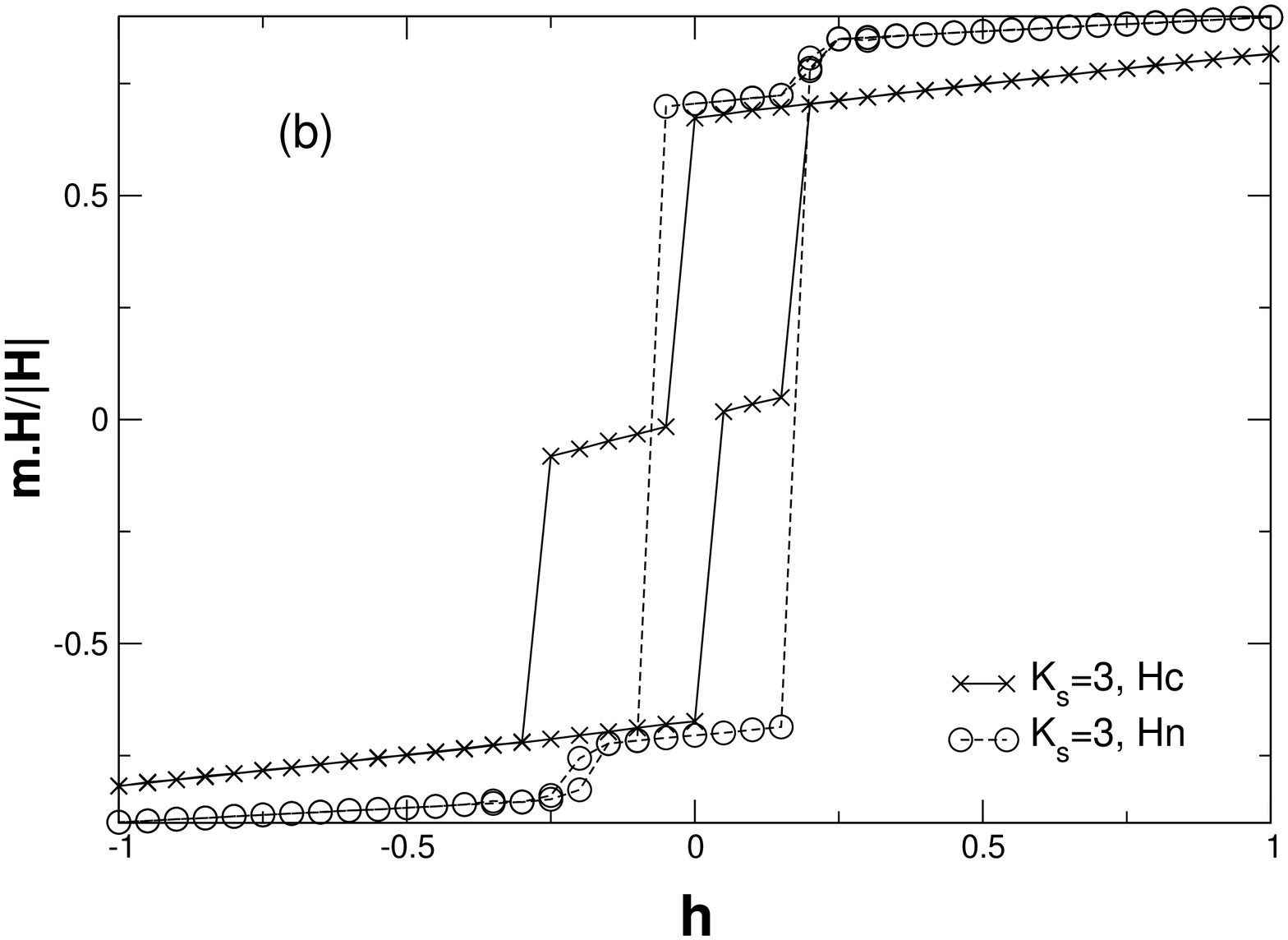}
\caption{Hcp lattice, hysteresis cycle, $T=0.01$.  $H_c$ denotes the magnetic field
 applied parallel to the $\vec u_c$ axis,
 $H_n$ denotes the magnetic field  parallel to the $\vec u_n$ axis (see figure~\ref{lattices}(a)).
Projection of the magnetisation along the axis of the corresponding applied field.
 The lines are only a guide to the eye. (a)  $K_{s}$=2,
 (b)  $K_{s}$=3.
 The cycle width is larger when the field is  parallel to the $\vec u_n$ axis.  
 In this case the magnetisation remains almost unchanged till $h=0$, showing intermediate jumps before
 saturating on the opposite sens.  When the  field is parallel to $\vec u_c$ a monotonous
  increase of the magnetisation with the applied field can be observed.}

\label{hyst_hcp}
\end{center}
\end{figure}

For the hcp lattice no hysteresis loop is observed for  $K_{s}$=1 but it is already 
present for  $K_{s}$=2.
We observe that the field orientation along $\vec u_c$ gives a smoother loop than
the orientation along $\vec u_n$ for all  $K_{s}$ values.   
In Figure~\ref{hyst_hcp}(a) it can be seen that
small plateaux, and a larger coercive field, appear when the field is applied 
along the $\vec u_n$ direction.

For $K_{s}$=3 these characteristics persist (see Figure~\ref{hyst_hcp}(b)).
Hysteresis disappears only in the $\vec u_c$ direction for  $K_{s}$=4
and for both directions for  $K_{s}$=5.

For the fcc lattice the dependence of the coercive field on the orientation of the applied field 
is the same as for hcp, but the cycle width starts diminishing already at  $K_{s}$=3.

For  these two lattices, the magnetisation shows small plateaux as a function of the applied field.
Step-like hysteresis cycles have been observed for large clusters with surface anisotropy
in~\cite{kach4,dim1,dim2} and for small antiferromagnetic 
clusters in~\cite{mc3}.
They have been explained in terms of the simultaneous reversal of a group of spins. Here we show,
additionally,  that the location of these plateaux depends not only on the intensity of the field but 
also on its direction with respect to the crystal axes of the clusters. 

For the icosahedral lattice the situation is once again different from that of the 
layered structures.  First, at
the temperature where we performed the previous field loops, 
the MC simulation fails to flip individual spins over the high energy barriers. 
This different temperature scale is again due to the higher coordination of the surface sites, 
which increases the effect of the ferromagnetic coupling. 
We have then performed
the field loops at $T=0.1$ (the hysteresis disappears for  $K_{s}$=3 at $T=0.15$).
Second, at a given $T$, the width of the hysteresis loops increases with  $K_{s}$, which indicates
that it is more difficult to reverse the magnetisation of the system  as  $K_{s}$ increases,
in contrast with the hcp and
fcc cases. Again, this can  be understood considering the peculiarities of this globular structure,
and comparing its ground state with those found for hcp and fcc (see 
figures~\ref{gs_conf_hcp},~\ref{gs_conf_fcc} and~\ref{gs_conf_ico}).
For a layered lattice one can see that there is a competition between the field and the 
anisotropy terms: when the cluster is (almost) completely polarized, for exemple, in the $+\vec u_c$
direction, the anisotropy
term in the hexagonal layer is far from being optimized (it is globally zero); 
when the field is then diminished in the reversal process,
 the anisotropy term ``helps'' the individual spin transitions. The situation is
different  for the icosahedral
structure in which the radial character of the anisotropy term allows for a configuration which
can follow the field without rising too much the anisotropy energy.  When the field is reversed 
at low $T$, again
in order to satisfy both terms, each individual spin must completely reverse its  orientation along 
its radial direction, so the intermediate states involve a jump over the anisotropy barrier, which 
increases with  $K_{s}$. This is exactly what we observe, the coercive field  increases 
with  $K_{s}$.  We thus believe that the single spin flip algorithm used is not adapted for the study of 
low temperature hysteresis loops for the icosahedral cluster.

\section{Conclusions}
\label{sec:4} 

We have studied the magnetic  thermal behaviour of a $N=13$ cluster in a magnetic field. 
We have considered a  classical Heisenberg model with  ferromagnetic interactions in a magnetic
field, and a  radial surface anisotropy term. 
This model has been applied to  three different lattice structures, hcp, fcc and icosahedral, 
which have similar cohesion energies in the context of Lennard-Jones and Gupta potentials.

Our results show that, even in zero field, the crystal structure of the cluster plays an
essential role on its thermal behaviour.
In particular, the anomalies seen in the average cluster magnetisation curves as a 
function of the temperature for some surface anistropy constants seems characteristic 
of the hcp structure.

This anomalous magnetic behaviour has been observed on $Gd_{13}$  clusters
~\cite{gerion} which are  found to present  a hcp structure~\cite{re4}.
Experimental measurements of the magnetic moment per atom of $Gd_{13}$ clusters
give values well below the one predicted by electronic structure
calculations for the ground state of the $Gd_{13}$~\cite{re4} and show 
a peak in the $m(T)$ curves~\cite{gerion,lopez}.

It has been proposed that a non-collinear arrangement of the atomic 
moments could be responsible for such a behaviour~\cite{gerion,re4}.
In ~\cite{re4} a simplified version of a RKKY model, with no anisotropy term, 
is proposed as a way to obtain
these non-collinear configurations.   
In fact, the oscillating character of the RKKY 
interactions has been replaced by a  
 ferromagnetic nearest-neighbour interaction and an antiferromagnetic coupling of  {\it all} 
 other pairs in the cluster, thus renforcing the competition. 
In~\cite{cerovski}, using the same  model on a hcp lattice, the authors find a range of values for 
the competing ratio between ferromagnetic and antiferromagnetic interactions, $\gamma$,
where the $m(T)$ curves in zero field have a peak. These two studies use a classical approach. 

On the other hand, in~\cite{lopez} a classical and a quantum study of the model has been carried out.  In this case they consider competing interactions: ferromagnetic between first neighbours and
 antiferromagnetic {\it only} between second neighbours.  They study N=13 atom closed packed clusters 
with: body
centered cubic, face centered cubic, hexagonal compact packing and icosahedral structures. They also 
study the N=14 hcp (open shell) cluster  with different geometries. For the closed packed (N=13)
 clusters, the results of
the  quantum approach are
  qualitatively the same as those of the   classical one found
 in~\cite{re4,cerovski}.  In all these studies the   behaviour of the $m(T)$ curves is the same: 
they show a  peak where the difference $\delta m= m(T_{max})-m(0)$ 
is less than $10\%$ of the maximum value ($m(T_{max}$).

In this paper, we show that the same qualitative behaviour may be induced 
by a simple hamiltonian including, in addition to the first neighbours ferromagnetic 
interaction, a radial surface anisotropy term which accounts for a  reduction of the 
symmetry of the crystal at the surface. 
Our results show that, for the hcp structure, the one which $Gd_{13}$ is assumed to 
have~\cite{re4}, the  competition  between the two energy terms leads to a peak 
in the $m(T)$ curves for a certain range of  $K_{s}$. For the same range of  $K_{s}$ 
the low temperature magnetisation is decreased and a non-collinear configuration appears, 
in qualitative agreement with the experimental results~\cite{gerion}. The $\delta m$ value we found is
of the same order than in the previous works.

In these two complementary approaches  the constants of the model
are over-estimated. In our case
the values of $ K_s$ leading to the peak are too big compared to the first neighbour interaction.
In the RKKY approach (classical or quantum) the   antiferromagnetic interaction 
intensity necessary to observe the peak is about $36 \%$ of the ferromagnetic one. 
This is  around one order of magnitude higher than the ratio 
between the first and the second peaks of the RKKY interaction.
In~\cite{cerovski} this is additionnaly
over-estimated by the fact that all the neighbours but the nearest are coupled 
antiferromagnetically with the same intensity.

In real RE clusters one can expect both effects (competing RKKY interactions 
and surface anisotropy) to be present.
They both contribute to the non-collinear order and this may allow for the peak in the $m(T)$ 
curves to be observed for more realistic values of $K_s$ and $\gamma$ than those used 
in all these works. It is also useful to notice that for small clusters the experimental estimations
 of Curie temperature are much bigger than for the bulk.  For instance, for $Gd_{13}$ it has been 
found $T_c>500K$ when bulk Gd Curie temperature is $T_{c}^{\it bulk}=293K$~\cite{gerion}.  This means that 
the cluster is able to maintain its magnetic order well beyond the temperature corresponding to 
the bulk. This pleads for a model which could be compatible with a canted structure and a 
stabilisation of magnetic order even for quite high temperatures. The antiferromagnetic second order
interaction in competition with the first neighbours one goes in the sense of  a lowering of the $T_c$.
On the contrary, the anisotropy term contributes to stabilize a canted magnetic structure.
This suggests that a more realistic model should include both therms in the hamiltonian.
 
In a Stern-Gerlach experiment, the projection of the magnetisation of {\it a
single particle}
along the field gradient direction is measured.
We have shown that this projection depends on the relative orientation 
between the applied field  and the crystal axes.
Then, different relative orientations, experimentally unknown, will 
broaden Stern-Gerlach deflection profiles. 
Such broadened deflection profiles have been observed for some RE
clusters~\cite{gerion}.
In this case, it is said that the clusters have a
{\it locked moment behaviour}~\cite{re1,gerion}.  
This means that the individual spins of the cluster are tightly coupled to the lattice by
the crystal anisotropies, finding it more difficult to follow
the applied field. This gives rise to broad deflection profiles.
We show additionally, that the relative orientation
of the applied field with respect to the crystal axes of the cluster
also contribute to the broadening of the deflection profiles.

We want to stress that, in this work, we deal with equilibrium properties of the 
clusters, so we are not describing the relaxation process taking place in 
 Stern-Gerlach apparatus.  This relaxation should  be considered 
in the interpretation of the results of such experiments.  
This has been done by an intermediate approach, besides the superparamagnetic model 
and locked-moment model, in both semiclassical~\cite{jensen} and quantum version~\cite{hamamoto}.
These  models consist in  treating the cluster as a single  magnetic 
moment coupled to the crystal by an uniaxial volume anisotropy term and to the applied field.
The rotation degrees of freedom allow for the cluster's magnetic moment  relaxation.
This  approach is complementary to ours. 
The anisotropy term is different and will not give the canted spin structure proposed for 
$Gd_{13}$~\cite{re4}. In addition, it does not describe the competition between
exchange and anisotropy energies since it does not allow for individual relaxation of the 
spins. So no dependence on the lattice structure can be obtained within this approach.

Hysteresis cycles at low temperature also show a dependence on the
lattice structure and surface anistropy as well as on the direction of the
applied field.

Our results show that the effect of the lattice structure on the magnetic behaviour 
cannot be neglected.
Moreover, as the crystalline structure is not known 
for general N (first principles results are available only for very small particules), 
it is not excluded that, at the temperatures of the experiment, different 
structures could be present for a given N.  Again, the different values of the low-temperature
magnetisation could broaden the deflection profile.

Most of our results are related to the fact that, for small clusters, the structural details
become important.  
Authors studying very large clusters focus mainly on the shape of the cluster which is cut out of 
a lattice having simple cubic  (or spinel) lattice structure.  In these studies, it is  shown
 how the surface anisotropy contributes to  the existence of steps in the hysteresis loops. 
 No anomaly in the $m(T)$ curves has been reported (see~\cite{kach1}).
Nevertheless, when N increases, the ratio of the number of 
spins at the surface to the total number of spins decreases and the second neigbours energy increases.
It is then interesting to consider a model taking into account at the same time competing interactions
and surface anisotropy to study  how these properties evolve with N.  This study is in progress.

\end{document}